\documentclass[lettersize,journal]{IEEEtran}
\usepackage{amsmath,amsfonts}
\usepackage{algorithmic}
\usepackage{algorithm}
\usepackage{array}
\usepackage[caption=false,font=normalsize,labelfont=sf,textfont=sf]{subfig}
\usepackage{textcomp}
\usepackage{stfloats}
\usepackage{url}
\usepackage{verbatim}
\usepackage{graphicx}
\usepackage{cite}
\usepackage{xcolor}
\newcommand*{\corr}[1]{\textcolor{black}{#1}}
\newcommand*{\rev}[1]{\textcolor{black}{#1}}

\begin{document}

\title{An Enhanced Intersecting-Lines Method for Optical Loss Characterization in SIS Heterodyne Receivers}

\author{Fran\c{c}ois Joint, Magnus Strandberg, Denis Meledin, Igor Lapkin, Sven-Erik Ferm, Mathias Fredrixon, Leif Helldner, Alexei Pavolotsky, Erik Sundin, Victor Belitsky, Vincent Desmaris
\thanks{F. Joint, M. Strandberg,  D. Meledin, I. Lapkin, A. Pavolotsky,  S.-E. .Ferm, M. Fredrixon, L. Helldner, E. Sundin, V. Belitsky, V. Desmaris are with the Department of Physics and Astronomy 
Chalmers University of Technology, Onsala Space Observatory 
SE-439 92 Onsala, Sweden
}
}
\maketitle

\begin{abstract}
Accurate knowledge of radio-frequency (RF) losses in front of superconductor--insulator--superconductor (SIS) mixers is essential for optimizing millimeter- and terahertz heterodyne receivers. However, such losses are difficult to extract from conventional hot/cold-load measurements because the measured receiver noise combines contributions from the input optics, the mixer, and the intermediate-frequency chain. In this work, we present a numerical implementation of the straight-line, or intersecting-lines, method for estimating input-referred RF noise and optical loss from standard Y-factor measurements. Hot/cold IF spectra are acquired while sweeping the local oscillator (LO) pumping level from under-pumped to over-pumped operation, so that each IF frequency bin gives a family of linear relations between receiver output power and input load temperature. We evaluate all pairwise intersections between these lines and analyze the crossing cloud to obtain the effective RF input noise. We repeat the analysis across the IF band and multiple LO tunings which provides a frequency-resolved characterisation of the input losses over the full band. Our implementation turns the straight-line technique into a reproducible framework for receiver characterisation and relative optical-loss comparison in low-loss front-end configurations.
\end{abstract}


\section{Introduction}
\IEEEPARstart{S}{uperconductor}--insulator--superconductor (SIS) mixers are among the most sensitive detectors for millimeter- and submillimeter-wave heterodyne receivers and are widely used in radio-astronomical instrumentation \cite{graauw_herschel-heterodyne_2010, meledin_sepia345_2022, belitsky_alma_2018, baryshev_alma_2015, mahieu_alma_2012}. As SIS mixer technology has matured, receiver noise temperatures have approached a few times the quantum limit over broad frequency ranges. In this regime, the total measured receiver noise is often no longer dominated only by the mixer itself, but also by losses in the radio-frequency (RF) input path, the quasi-optical coupling system, cryostat windows, infrared filters, beam splitters, diplexers, and other passive components placed in front of the mixer. Even small losses before the mixer are particularly detrimental because they attenuate the input RF signal before detection while adding thermal noise determined by the physical temperature of the lossy element.

The standard hot/cold-load, or Y-factor, measurement provides the overall receiver noise temperature for a given operating point. However, this scalar measurement does not directly separate the contributions from the RF input section, the SIS mixer, and the intermediate-frequency (IF) chain. This separation is essential for receiver optimisation: RF input loss indicates limitations in the optical coupling or front-end components, IF noise identifies limitations after the mixer, and the remaining contribution is associated with the intrinsic mixer performance. Direct component-by-component loss measurements at millimeter and terahertz frequencies are often difficult, especially when the relevant losses are only a few percent and occur in cryogenic or quasi-optical environments \cite{filinis_development_2026}. A practical in-situ method that extracts RF input loss information from standard receiver measurements is then highly valuable.

The straight-line, or intersecting-lines, method uses standard hot/cold-load measurements at several reduced LO drive levels to separate noise contributions in SIS receivers~\cite{blundell_understanding_1992}. The method relies on the fact that changing the LO drive strongly changes the mixer conversion loss, but the mixer output noise can remain approximately constant in a suitable low-pumping range. Each LO setting then creates a straight line in the IF-output-power versus load-temperature plane. If those lines share a common crossing point, its horizontal coordinate gives an equivalent input noise temperature, and the vertical coordinate is related to the IF-chain contribution. Ke and Feldman later showed, using quantum mixer theory that under the same low-pumping assumption this input noise corresponds to the RF input section~\cite{ke_technique_1994}. The straight line method gives access to RF/input loss without calibrated IF components or loss model of every optical element.

The method was later reformulated in a more experimentally convenient form, with a direct estimate of the uncertainty on the intersecting temperature~\cite{tong_empirical_2008, tong_quasi-optical_2020}. Instead of constructing the intersection in the output-power plane, the receiver noise temperature is plotted as a function of relative conversion loss. In the low-LO-drive regime, the data follow a straight line, and the extrapolated intercept gives the intersecting temperature (denoted $T_X$). This has two practical advantages. First, deviations from linearity provide a direct indication of the LO-drive range over which the assumptions of the method remain valid. Second, the scatter around the fitted line gives an estimate of the experimental uncertainty. Tong \emph{et al.} also showed that, when measurements are repeated with and without an additional optical component, the change in $T_X$ can be used to estimate the insertion loss of that component.

More recent work has revisited the physical basis of the method for practical SIS mixers, showing that the low-LO receiver noise can be written in a form that separates the RF input noise from gain-dependent mixer and IF-chain contributions~\cite{shan_inconstant_2024}. In this limit,
\begin{equation}
T_{\mathrm{RX}}= T_{\mathrm{RF}}+T_1+\frac{T_0+T_{\mathrm{IF}}}{G_{\mathrm{mix}}},
\end{equation}
where $T_{\mathrm{RF}}$ is the input noise associated with the RF section, $T_1$ represents the part of the mixer noise which remains approximately constant when referred to the mixer input, and $T_0$ and $T_{\mathrm{IF}}$ are additional output-noise contributions from the mixer and IF chain, respectively. This explains why the intersecting-line method can be applied only in the low-LO range. If changing the LO pump changes the mixer conversion gain $G_{\mathrm{mix}}$, the RF input contribution should stay as a unique crossing point. With these conditions, a family of hot--cold load lines can intersect near a common input temperature, which allows us to determine the RF input noise from the crossing point.

In this work, we build on the straight-line and intersecting-lines methods by developing an enhanced numerical implementation suitable for \corr{heterodyne receivers based on SIS mixers}. Instead of applying the method to a single detector bandwidth or to a small number of LO-drive settings, we analyse complete hot/cold IF spectra over many LO powers and multiple LO frequencies. For each LO frequency and each IF-frequency bin, the measured hot and cold powers create a family of straight lines in the output-power--versus--load-temperature plane. The crossing point is estimated numerically from the full set of lines, which gives a frequency-dependent RF input noise temperature. This allows us to resolve spectral structure in the extracted noise, including standing waves, weak resonances, and frequency-dependent optical-coupling effects that would otherwise be hidden in a Y-factor analysis.

The use of many LO-drive points also makes the crossing problem overdetermined, which allows us to estimate the uncertainty of the extracted input noise temperature from the distribution of pairwise intersections. These provide a diagnostic of the validity of the straight-line assumptions and help us to identify frequency regions or LO-drive ranges affected by mixer-output-noise variation, bias-point changes, or optical reflections. We then used the method in a relative configuration to determine the loss introduced by a transparent medium placed in the receiver input path. In this case, the change in the extracted RF input noise between the reference and sample configurations is used to determine the added optical loss \cite{shan_inconstant_2024, tong_quasi-optical_2020}.

\section{Method}
\subsection{Receiver Under Test}
\begin{figure}[!t]
\centering
\includegraphics[width=3.5in]{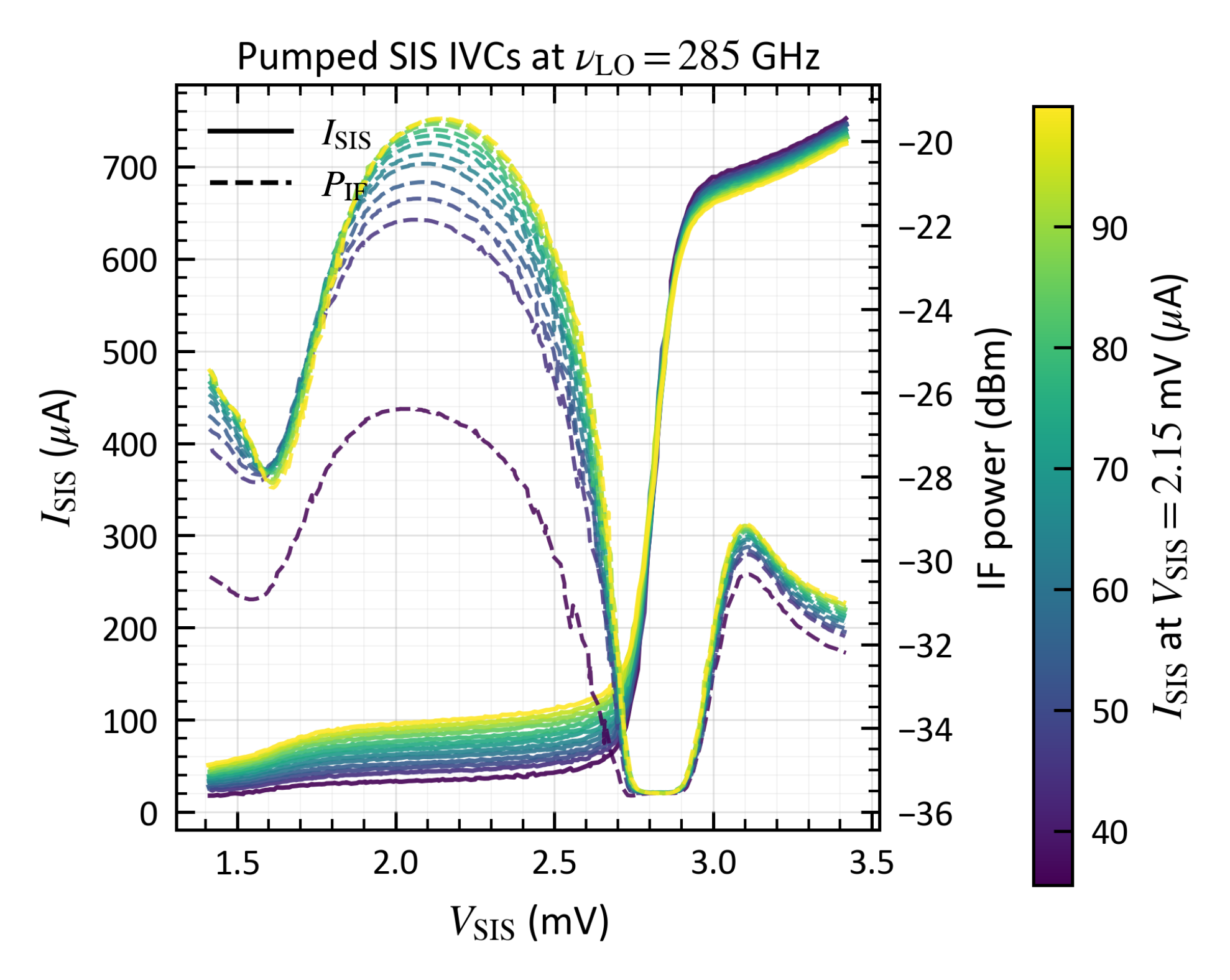}
\caption{LO-pumped characterization of the SIS heterodyne mixer used in this work. The figure shows the SIS current, $I_{\mathrm{SIS}}$, and the IF output power, $P_{\mathrm{IF}}$, as functions of the SIS bias voltage, $V_{\mathrm{SIS}}$, for different LO pump levels. The pump level is indexed by the SIS current measured near the center of the first photon-assisted tunneling step. These define the LO-drive range used for the straight-line analysis. \corr{All curves are recorded with the 295 K load at the mixer input.}}
    \label{fig:IVC}
\end{figure}
The receiver used in this work is a double-sideband SIS mixer that covers approximately 220--375 GHz, with a wide IF bandwidth of 4--20 GHz \cite{lapkin2025Dev, lapkin2026Dev}. At the time of the measurements, the mixer was still under development and had not yet been fully optimized as a final receiver cartridge. In this work, it was used as a stable experimental platform to validate the straight-line analysis over a large number of LO and IF operating points. The measurements focus on the 275--375 GHz range, corresponding to the ALMA Band 7 frequency window. \corr{The receiver uses an RF quadrature hybrid, with one output feeding the DSB SIS mixer used in this work and the other terminated with a 50~$\Omega$ waveguide load \cite{lopez_microfabricated_2025}. The mixer output was amplified by a cryogenic Yebes HEMT low-noise amplifier~\cite{lopez-fernandez_16-ghz_2024} followed by a room-temperature Low Noise Factory amplifier, providing approximately 60~dB of total IF gain over 4--20~GHz.}

\corr{The LO used is a room-temperature active multiplication chain (AMC) of total factor $\times$18. A microwave synthesiser delivered 9\,dBm over 15.8--20.3\,GHz into a $\times 6$ multiplier chain followed by a power amplifier, which 95--118\,GHz output drove a WR2.8 $\times 3$ frequency tripler so that the chain could cover 285--365\,GHz. The LO power is brought into the cryostat through an oversized waveguide, with transitions at the cryostat feedthrough and at the mixer-block LO input port, and the pump level is set by the drive power of the chain, which gives the nine junction currents $I_{\mathrm{SIS}} \approx 40$--90\,$\mu$A used throughout.}

The optical path in front of the mixer consists of a vacuum window \cite{joint_3d-structured_2026}, infrared-blocking filters mounted at the 50~K and 15~K stages, a pair of off-axis parabolic mirrors, a corrugated horn antenna and an orthomode transducer.

\begin{figure}[!t]
\centering
\includegraphics[width=3.45in]{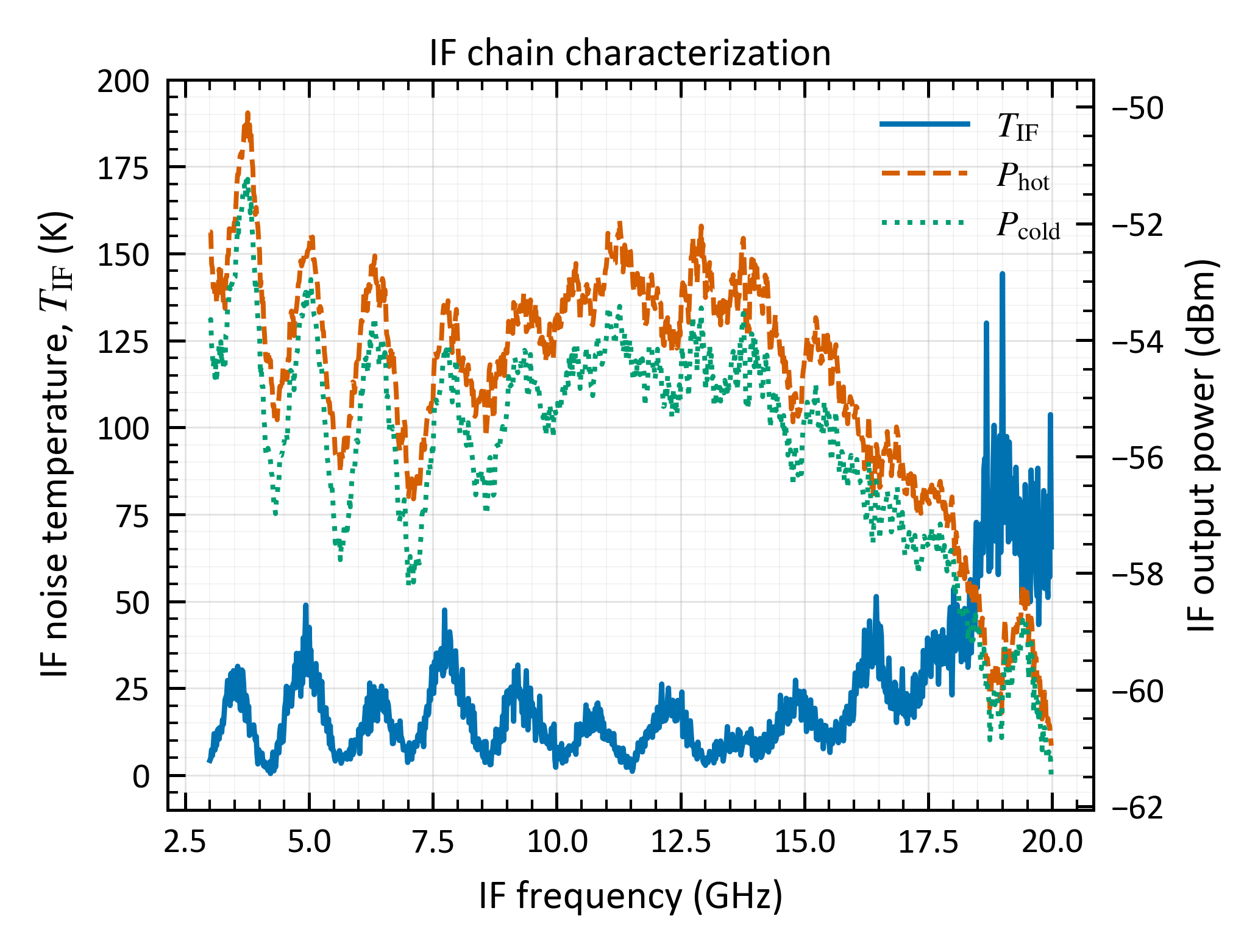}
\caption{IF-chain characterization used in the receiver-noise analysis. The IF noise temperature and the corresponding IF output power are plotted as functions of IF frequency. These characterize the frequency-dependent IF contribution and provide a reference for interpreting the hot/cold receiver spectra used in the straight-line analysis.}
    \label{fig:T_IF}
\end{figure}

\corr{Figs.~\ref{fig:IVC} and \ref{fig:T_IF}} show representative diagnostic measurements of the mixer. The dc current--voltage characteristic of the twin SIS junctions is plotted near the superconducting gap, together with the detected IF output power of the pumped SIS junction. Under LO illumination (\rev{285~GHz}), a photon-assisted tunneling step is observed in the pumped I--V curve. In the same bias sweep, the IF output power measured with a power meter shows the corresponding LO-induced response as a dome-like feature around the pumped photon-assisted step. Throughout this work, the LO pumping level is referenced by the SIS current measured at the centre of this photon-assisted step.

As an additional diagnostic of the IF chain, the mixer was biased on the normal-state branch of its current-voltage characteristic and used as a calibrated shot-noise source \cite{belitsky_shot_1990}. IF output spectra were recorded at two different normal-state bias voltages, $V_1$ and $V_2$j. The generated shot-noise temperature of a tunnel junction can be written, in the usual high-bias form, as

\begin{equation}
    T_{\mathrm{shot}}(V) \simeq \frac{eV}{2k_{\mathrm{B}}}
\end{equation}

For each IF frequency, the measured output power can then be expressed as

\begin{equation}  
    P_{\mathrm{IF}}(V,\nu_{\mathrm{IF}}) =
    k_{\mathrm{B}} B G_{\mathrm{IF}}(\nu_{\mathrm{IF}})
    \left[
    T_{\mathrm{shot}}(V) + T_{\mathrm{IF}}(\nu_{\mathrm{IF}})
    \right]
\end{equation}

where B is the detection bandwidth, $G_{\mathrm{IF}}$ is the IF gain, and $T_{\mathrm{IF}}$ is the noise temperature of the IF backend. The two bias points therefore allow $T_{\mathrm{IF}}(\nu_{\mathrm{IF}})$ to be estimated without requiring heterodyne conversion.

The resulting IF noise-temperature spectrum shows standing-wave structure across the band. This is expected in this configuration, since the SIS junction is biased
in its normal state and is not matched to the 50\,$\Omega$ amplifier input. \corr{The
shot-noise calibration was not corrected for this mismatch, which biases the
inferred $T_{\mathrm{IF}}$ low. Over 4--16\,GHz the mean $T_{\mathrm{IF}}$ is
14~K across the band; the points approaching zero occur where the mismatch is largest and are not physical. The measured spectra nevertheless confirm that the backend remains operational over the frequency range relevant to the present analysis, with useful performance up to approximately 16~GHz.}

\subsection{Numerical implementation}
\begin{figure}[!t]
\centering
\includegraphics[width=3.45in]{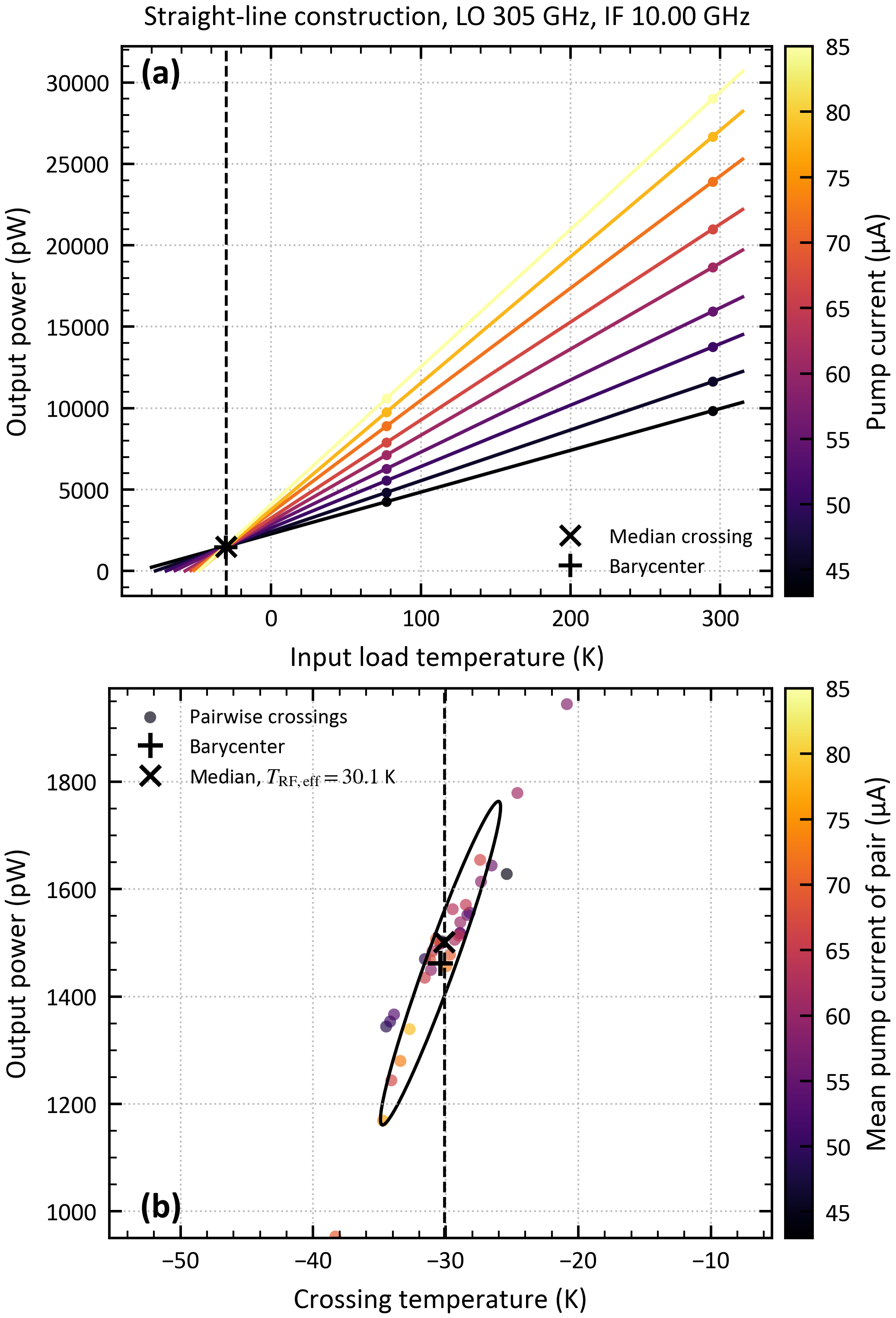}
\caption{\corr{Straight-line analysis at $\nu_{\mathrm{LO}}=305~\mathrm{GHz}$ and $\nu_{\mathrm{IF}}=10~\mathrm{GHz}$. }(a) IF output power as a function of load temperature for several LO pump levels, showing the family of straight lines constructed from the hot/cold measurements. (b) \corr{Pairwise crossing points of these lines, together with the confidence ellipse and the corresponding weighted centroid and median.} The effective input-referred noise temperature ($T_{\mathrm{RF}}$) is inferred from the horizontal coordinate of the common crossing.}
    \label{fig:StraightLine}
\end{figure}
To characterize the receiver over the full band, we recorded standard hot--cold load measurements of the receiver IF spectrum using room-temperature and liquid-nitrogen loads, with physical temperatures of 295~K and 77~K, respectively. \rev{For the analysis, these physical temperatures were converted to Callen–Welton radiation temperatures.} For each LO frequency, the LO drive was swept over nine pumping settings, referenced to the SIS current measured at the centre of the photon-assisted step in the pumped I--V curve. The current range extended from 40 to 90~$\mu$A, from under-pumped, near-optimum, to over-pumped operating conditions; the optimum pumping level for this mixer was around 75~$\mu$A. This pump sweep was deliberately chosen in order to test the consistency of the straight-line assumption and to identify pump ranges where the crossing point becomes poorly defined.

The measurement was repeated for 15 LO frequencies spaced by 5 GHz across the receiver band. For each LO and pump setting, the IF output spectrum was recorded from 3 to 20~GHz with a frequency step of approximately 17~MHz, giving 1001 IF points per spectrum. The complete dataset consists of hot and cold IF spectra sampled over LO frequency, LO pumping level, and IF frequency. \corr{A complete set for one receiver, 15 LO settings $\times$~9 pump settings  $\times$~2 loads, takes approximately 90 min, dominated by the LO retuning.}
\begin{figure}[!t]
\centering
\includegraphics[width=3.45in]{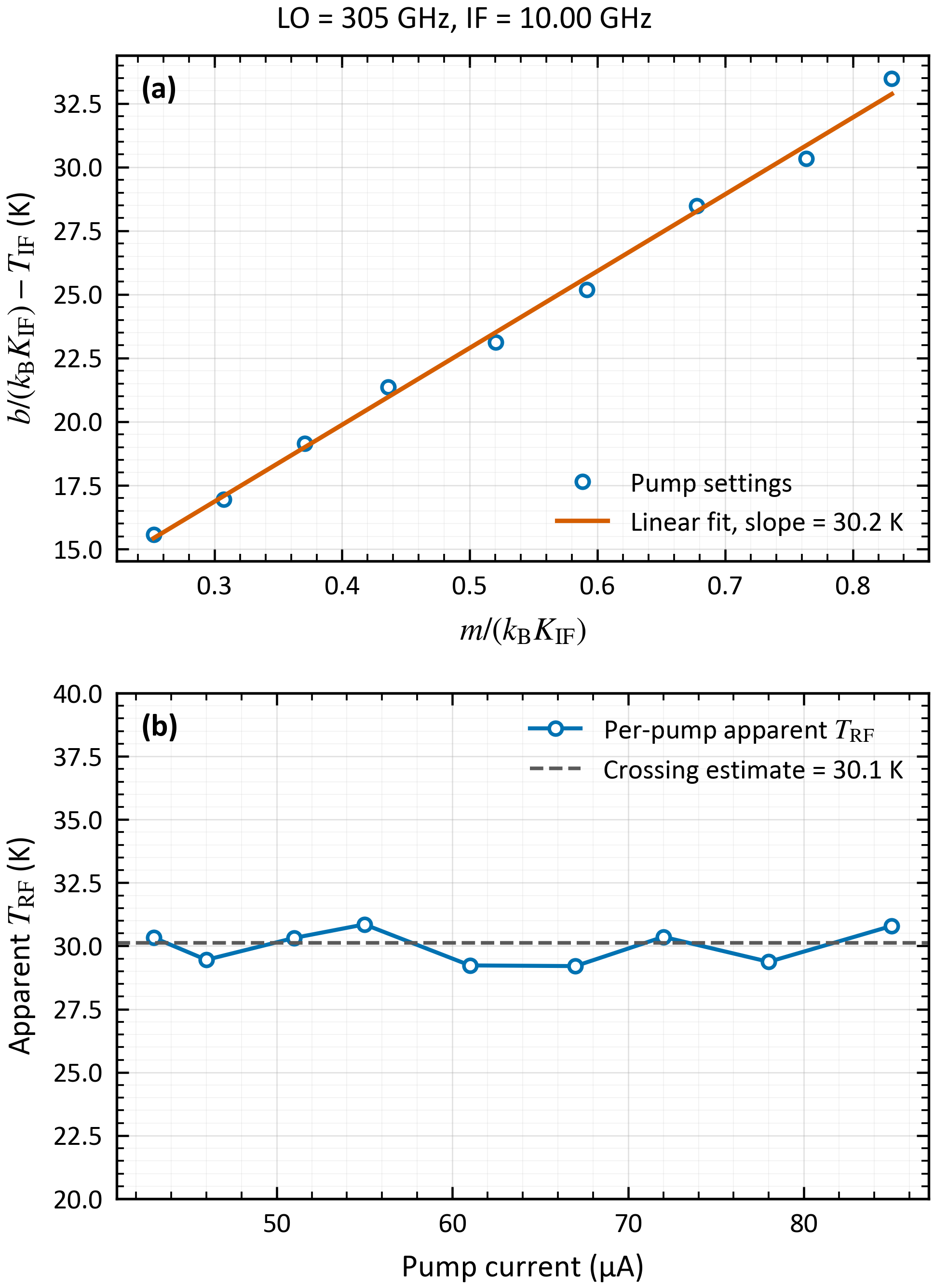}
\caption{\corr{Empirical consistency check of the straight-line extraction at $\nu_\mathrm{LO}=305$~GHz and $\nu_\mathrm{IF}=10.0$~GHz.}
(a) Calibrated intercept versus measured conversion product for different LO pump level; the fit slope gives the effective RF noise temperature.
(b) Apparent $T_\mathrm{RF}$ obtained from each pump line using the crossing ordinate $P^\ast$, compared with the crossing-derived $T_\mathrm{RF,eff}$ dashed line.
}
    \label{fig:theory_alt}
\end{figure}

For a fixed LO frequency and IF bin, each LO pumping setting defines one straight line in the plane of IF output power versus input load temperature (Fig.~\ref{fig:StraightLine}a). The line is constructed from the two hot- and cold-load measurements as $P_i(T)=m_iT+b_i$
where $i$ denotes the LO pumping setting. The slope $m_i$ is proportional to the effective conversion from input temperature to detected IF power, and therefore changes strongly with LO pumping. The intercept $b_i$ contains additive terms that are weakly dependent on the input load temperature, including the IF-chain contribution~\cite{honingh1997determining}.

Instead of determining a single crossing point, we compute all pairwise intersections between the lines. For $N=9$ pumping settings this gives $N_{\mathrm{cross}}=N(N-1)/2=36$ crossing points for each LO--IF bin before rejection of ill-conditioned pairs (Fig.~\ref{fig:StraightLine}b). For two lines $i$ and $j$, the crossing coordinates are
\begin{equation}
T^{*}_{ij}=\frac{b_j-b_i}{m_i-m_j},
\qquad
P^{*}_{ij}=m_iT^{*}_{ij}+b_i .
\end{equation}
The two crossing coordinates have a direct physical meaning. Writing the response of pump setting $i$ as a line through $(T^{*},P^{*})$, the input temperature at which the extrapolated line reaches zero output power gives the receiver noise temperature,
\begin{equation}
T_{\mathrm{noise}}^{\mathrm{DSB}}
= -T^{*}+\frac{P^{*}}{m_i}
= T_{\mathrm{RF}}+\frac{P^{*}}{m_i},
\label{eq:tnoise}
\end{equation}
\rev{where $T_{\mathrm{RF}}=-T^{*}$ represents the effective input-reffered noise contribution extracted from the crossing temperature, and $P^{*}/m_i$ represents the contribution associated with the mixer output and IF-chain noise, referred to the receiver input. The use of Callen–Welton radiation temperatures incorporates the zero-point contribution $h\nu/2k_B\approx$7.2~K at 300 GHz into the calibration loads.} \rev{This contribution is therefore already accounted for in the incident radiation and is not added separately to the extracted receiver noise.} This pairwise construction preserves the full crossing distribution instead of reducing the data immediately to a single number. For each crossing cloud we calculate both the coordinate-wise median and the \rev{arithmetic centroid of the retained pairwise intersections. The median provides a robust estimate of the crossing temperature by reducing the influence of extreme intersections, particularly those arising from nearly parallel lines. The centroid is used only for visualization, together with the confidence ellipse.} Such effects as nearly parallel lines, residual standing waves, pump settings outside the valid low-LO regime, or deviations from the weakly varying mixer-noise assumption would appear as a broadened or skewed crossing cloud. The resulting distribution gives us an estimate of the effective crossing temperature and a diagnostic of the validity of the extraction. The weighted centroid is useful as a visual estimate of the centre of the distribution and is shown in Fig.~\ref{fig:StraightLine}b with the confidence ellipse. Although, unless otherwise stated, the value used for the extracted input-referred temperature is the median crossing temperature, $T_{\mathrm{RF}} = -\mathrm{median}\left(T^{*}_{ij}\right)$. 

\subsection{Calibrated analysis}
To estimate the validity of the straight-line extraction, we place the measured hot--cold slopes on an absolute gain scale using the IF-chain gain--bandwidth product $K_\mathrm{IF}=G_\mathrm{IF}B$, obtained from the shot-noise measurement described above (Fig.~\ref{fig:T_IF}). In that calibration, the two normal-state SIS bias points, $V_\mathrm{SIS}=4$ and 6~mV, correspond to equivalent shot-noise temperatures of 23.2 and 34.8 K, respectively, which gives $K_\mathrm{IF}=\frac{P_\mathrm{IF}(6~\mathrm{mV})-P_\mathrm{IF}(4~\mathrm{mV})} {k_\mathrm{B}(34.8-23.2)}.$ The same predetection bandwidth $B$ is used for the shot-noise calibration and for the receiver hot--cold measurements. For each LO pump setting $i$, the measured hot--cold slope is therefore $m_i=k_\mathrm{B}B\,G_\mathrm{RF}G_{\mathrm{mix},i}G_\mathrm{IF}$ so that the corresponding total system gain is $G_{\mathrm{sys},i}=\frac{m_i}{k_\mathrm{B}B}=G_\mathrm{RF}G_{\mathrm{mix},i}G_\mathrm{IF}$. Dividing by the independently measured IF gain, or equivalently using $K_\mathrm{IF}=G_\mathrm{IF}B$, gives $\frac{m_i}{k_\mathrm{B}K_\mathrm{IF}}=G_\mathrm{RF}G_{\mathrm{mix},i},$ provided that the RF transmission remains unchanged during the LO pump sweep. With the same normalization, the calibrated intercept $b_i/(k_\mathrm{B}K_\mathrm{IF})$ against the calibrated slope $m_i/(k_\mathrm{B}K_\mathrm{IF})$

\begin{equation}
\frac{b_i}{k_\mathrm{B}K_\mathrm{IF}}=
    G_\mathrm{RF}G_{\mathrm{mix},i}T_\mathrm{RF}
    + T_1G_{\mathrm{mix},i} + T_0 + T_{\mathrm{IF}}
\end{equation}
Therefore,
\begin{equation}
    \frac{b_i}{k_\mathrm{B}K_\mathrm{IF}}=
    \left(T_\mathrm{RF}+\frac{T_1}{G_\mathrm{RF}}\right)
    \frac{m_i}{k_\mathrm{B}K_\mathrm{IF}}
    + T_0 + T_{\mathrm{IF}}
\end{equation}
A plot of the calibrated intercept $b_j/(k_\mathrm{B}K_\mathrm{IF})$ against the calibrated slope $m_j/(k_\mathrm{B}K_\mathrm{IF})$ should therefore be linear if $T_0$, $T_1$, and $T_{\mathrm{IF}}$ remain approximately constant over the selected low-pumping range. The fitted slope gives an effective input-referred term,
\begin{equation}
    T_{\mathrm{RF,eff}}=
    T_\mathrm{RF}+\frac{T_1}{G_\mathrm{RF}}
\end{equation}
which reduces to the RF input noise only when the gain-dependent mixer-noise contribution $T_1$ is negligible, or when it is independently accounted for. Deviations from linearity provide a direct check of the main assumption behind the intersecting-line method~\cite{honingh1997determining ,shan_inconstant_2024}.

\rev{Fig.~\ref{fig:theory_alt}a} shows the calibrated intercept as a function of the measured gain product for the different pump settings. The linear dependence indicates that the pump-dependent change in intercept is mainly explained by the change in conversion gain, and not by a strong pump dependence of the additive output-noise term. As a complementary check, Fig.~\ref{fig:theory_alt}b shows the apparent RF noise temperature obtained from each individual pump line using the common crossing ordinate $P^\ast$,
\begin{equation}
T_{\mathrm{RF},i}^{\mathrm{app}}=\frac{b_i-P^{\ast}}{m_i}.
\end{equation}
\rev{The low spread of $T_{\mathrm{RF},j}^{app}$ in Fig.~\ref{fig:theory_alt}b around the crossing-estimate value shows that the extracted $T_{\mathrm{RF}}$ is not strongly dependent on the selected pump state.}

Although this analysis is not a full quantum-theory mixer analysis, as $G_\mathrm{mix}$ and $T_{\mathrm{out},j}$ are not  calculated independently, it  tests that pump-dependent output noise or other additive terms could shift the crossing point and lead to an overestimate of $T_{\mathrm{RF}}$. The linearity of the calibrated intercept and the limited spread of $T_{\mathrm{RF},j}^{app}$ support the robustness of the extracted $T_{\mathrm{RF}}$ over the selected operating range.

\begin{figure}[!t]
\centering
\includegraphics[width=3.45in]{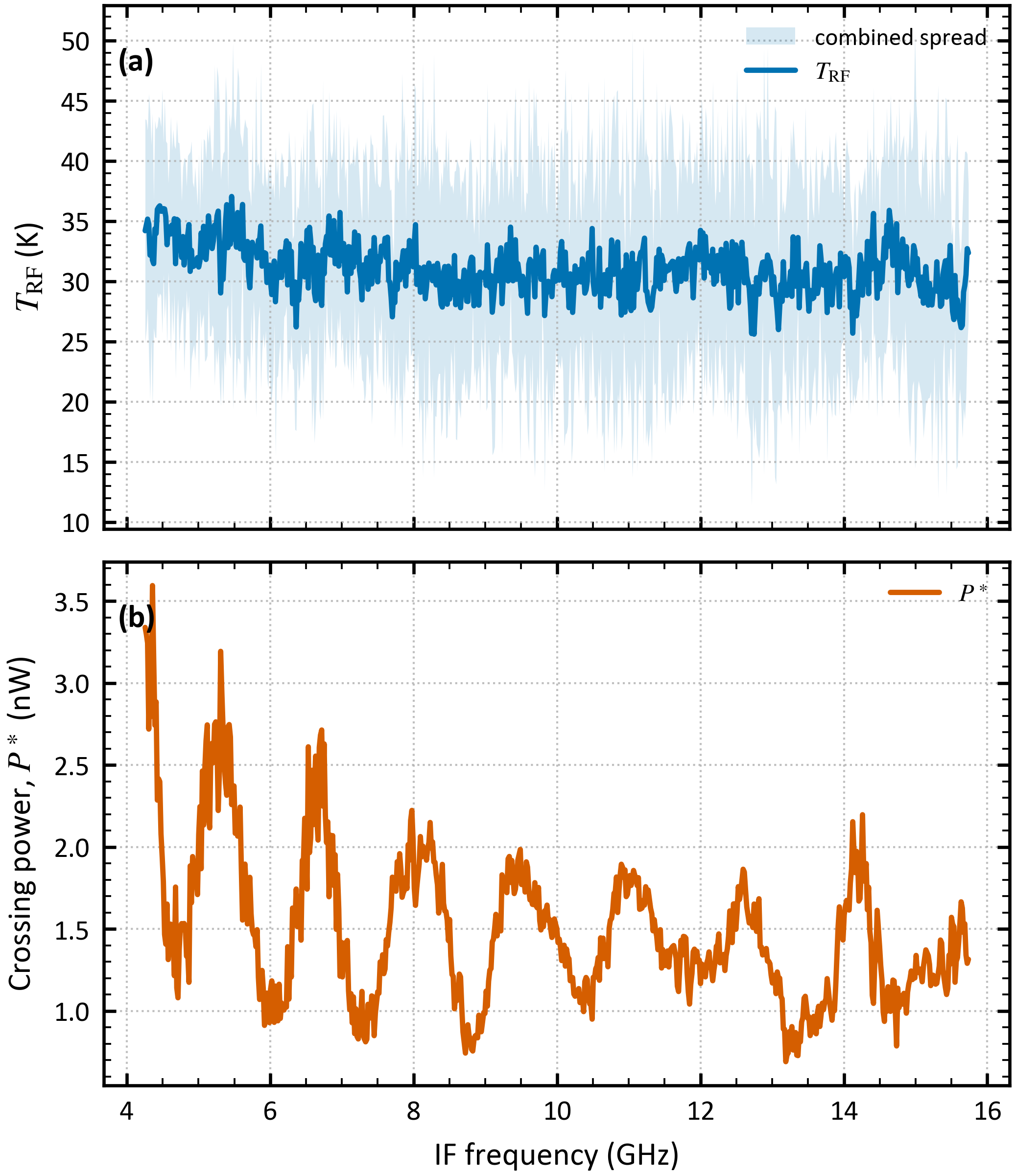}
\caption{\corr{IF-resolved straight-line crossing analysis for a LO at 305~GHz.} (a) Extracted effective input-referred temperature \corr{$T_{\mathrm{RF}}$} as a function of IF frequency. The solid line gives the crossing estimate, while the shaded band indicates the confidence interval obtained from the spread of the pairwise crossing points. (b) Corresponding crossing output power $P^\ast$, extracted from the same crossing analysis. The simultaneous variation of $T_{\mathrm{RF}}$ and $P^\ast$ across the IF band provides a diagnostic of the consistency of the crossing solution and of possible IF-dependent contributions to the measured receiver output.}
\label{fig:IF_resolved}
\end{figure}
\subsection{RF input noise temperature}
For the optical-loss analysis, the same crossing and calibration checks are repeated independently for every LO frequency and IF bin. 

\begin{figure}[!t]
\centering
\includegraphics[width=3.45in]{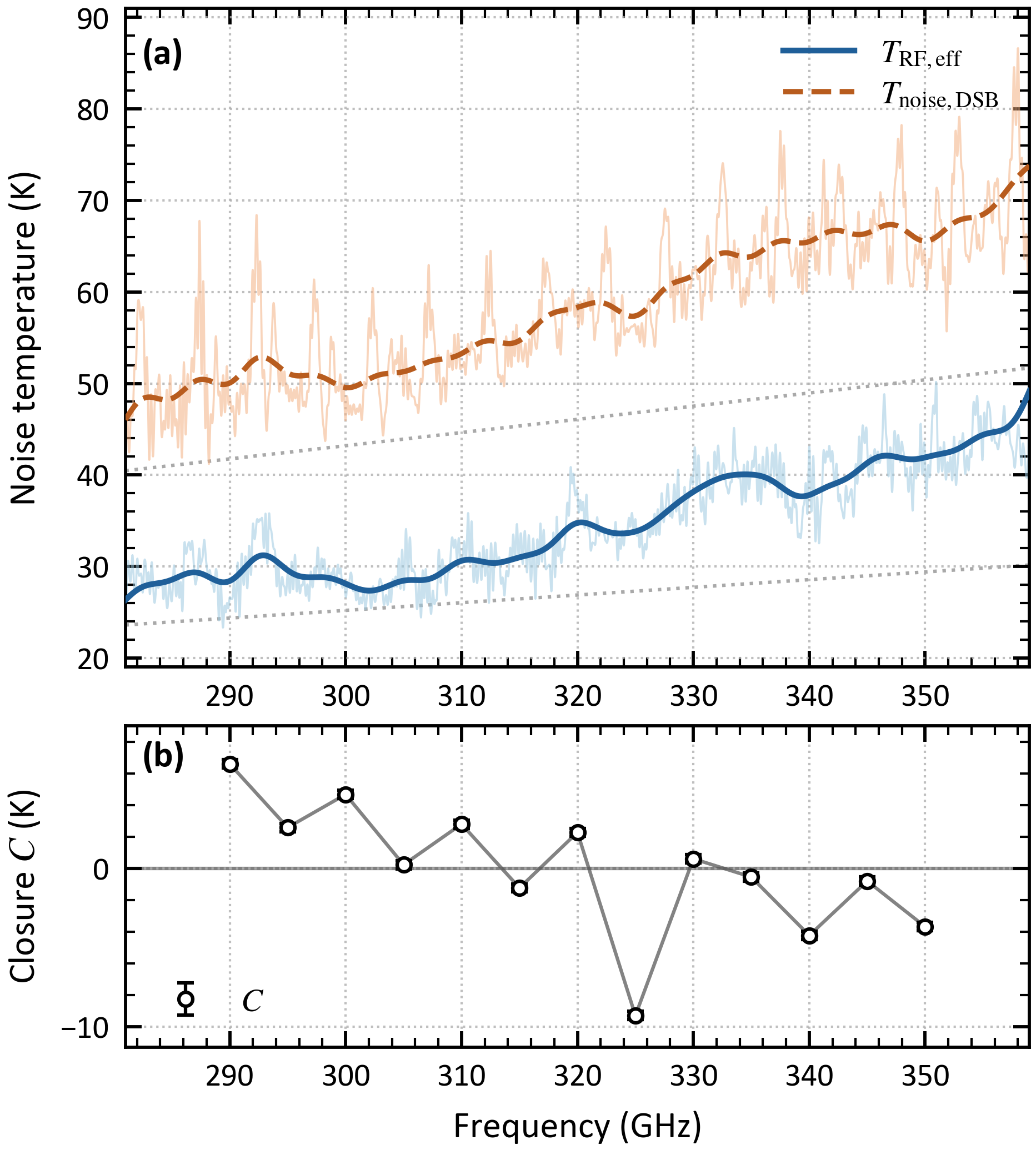}
\caption{(a) Sky-frequency spectrum of the input RF noise temperature, $T_{\mathrm{RF}}$, obtained from the DSB stitching of the IF-resolved straight-line estimates. The measured DSB receiver noise temperature, $T_{\mathrm{noise,DSB}}$, is overlaid for comparison.
(b) Adjacent-LO closure $C$, which probes residual tuning-dependent structure.
}
\label{fig:trf_tnoise_closure}
\end{figure}
For each LO--IF point, the spread of the extracted value is first estimated from the pairwise crossing temperatures $T^{*}_{ij}$. To plot a frequency-resolved result, we make a combined spread from the overlap of all estimates contributing to each IF bin. The central curve is the median extracted value, and the shaded region represents the resulting overlap spread (Fig.~\ref{fig:IF_resolved}a). At $\nu_{\mathrm{LO}}=305$~GHz and for the 4--16~GHz IF range, the extracted $T_{\mathrm{RF}}$ is between 30 and 35~K, with an overlap spread of approximately $\pm 8$~K. The same crossing analysis also yields the vertical coordinate $P^{*}$ (Fig.~\ref{fig:IF_resolved}b). $P^{*}$ is in the nW level and shows a standing-wave pattern versus IF frequency which is consistent with the measured IF noise temperature using the shot noise method.

\corr{To obtain a band-wide view of the input noise (Fig.~\ref{fig:trf_tnoise_closure}), we combine all LO tunings in a single reconstruction. Each crossing provides one double-sideband value, which is related to the two corresponding sky frequencies by
\begin{equation}
T_{\mathrm{RF}}^{\mathrm{DSB}}(\nu_{\mathrm{LO}},\nu_{\mathrm{IF}})
=
\frac{1}{2}T_{\mathrm{RF}}(\nu_{\mathrm{LO}}+\nu_{\mathrm{IF}})
+
\frac{1}{2}T_{\mathrm{RF}}(\nu_{\mathrm{LO}}-\nu_{\mathrm{IF}}),
\label{eq:dsb}
\end{equation}
where equal upper- and lower-sideband weighting is assumed. Because adjacent LO tunings overlap over the analysed IF band, the same sky frequencies are constrained repeatedly with different sideband combinations. Instead of assigning each measured value independently to both sidebands, we solve all equations simultaneously. Writing the system as $\mathbf{y}=\mathbf{A}\mathbf{x}$, where $\mathbf{y}$ is the measured DSB temperatures, $\mathbf{x}$ is the unknown sky-frequency spectrum, and $\mathbf{A}$ maps each
$(\nu_{\mathrm{LO}},\nu_{\mathrm{IF}})$ measurement on the upper- and lower-sideband frequencies, the spectrum is refined iteratively as
\begin{equation}
\mathbf{x}^{(k+1)}
=
\mathbf{x}^{(k)}
+
\alpha\mathbf{A}^{\mathsf T}
\left(\mathbf{y}-\mathbf{A}\mathbf{x}^{(k)}\right),
\label{eq:reconstruction}
\end{equation}
where $\mathbf{A}^{\mathsf T}$ back-projects the residual on the sky frequencies which produced it and $\alpha$ is a normalisation factor. In the numerical method, the update is normalised by the measurement redundancy. Only sky-frequency bins constrained by at least three independent LO tunings are retained. The same reconstruction is applied to the corresponding receiver noise, $T_{\mathrm{noise}}^{\mathrm{DSB}}=T_{\mathrm{RF}}+P^{*}/m$, which provides $T_{\mathrm{RF}}$ and $T_{\mathrm{noise}}$ on the same wideband sky-frequency grid.}
\corr{The consistency of the stitched solution was also tested using the closure quantity shown in Fig.~\ref{fig:trf_tnoise_closure}b. The closure compares two independent combinations of adjacent LO tunings which sample the same sky frequencies. In the ideal picture, the closure should be zero. The measured closure has an rms value of about 4.1~K, with the largest deviation occurring near an LO frequency of 325~GHz, where $C\simeq -9.7$~K (uncertainty estimate 0.26~K). This departure shows that part of the measured $T_{RF}$ \rev{can not} be represented only by a sky-frequency-dependent input \rev{noise. This} is consistent with an additional tuning-dependent contribution, such as LO AM noise, residual standing-wave effects, or other pump-dependent receiver fluctuations.}

\subsection{Extraction of a Room-Temperature Window Transmission}
\begin{figure}[!t]
\centering
\includegraphics[width=3.45in]{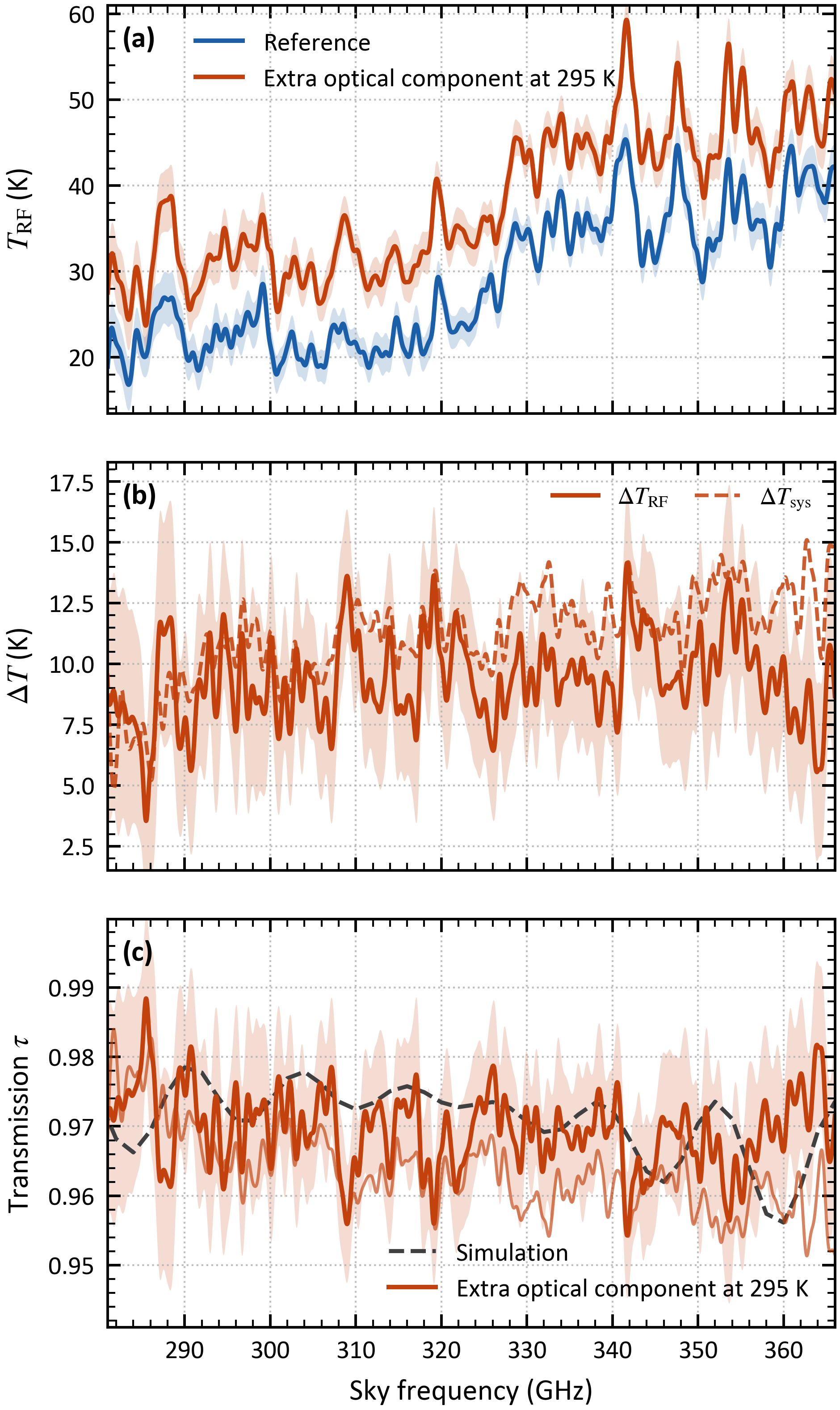}
\caption{Extraction of the room-temperature window transmission using the straight-line method.
(a) Extracted  $T_\mathrm{RF}$ as a function of sky frequency for the reference configuration and with the additional window inserted.
(b) Added input-referred noise $\Delta T_\mathrm{RF}$ obtained from the difference between the two configurations. 
(c) Window transmission derived from the measured change in $T_\mathrm{RF}$ using $\tau = (T_{\mathrm{RF,ref}} + T_\mathrm{phys})/(T_{\mathrm{RF,win}} + T_\mathrm{phys})$, with $T_\mathrm{phys}=295$ K. The dashed line shows the FEM transmission model. Shaded regions indicate the propagated spread of the extracted $T_\mathrm{RF}$ values. \corr{The dashed curve in (b) is the difference taken on the receiver
noise, $\Delta T_{\mathrm{sys}}$, and the thin curve in (c) is the transmission
obtained by inverting it.}}
\label{fig:window_transmission}
\end{figure}

As an independent application of the numerical straight-line analysis, the receiver was measured in two optical configurations: first in the reference configuration, and then with an additional room-temperature transparent window inserted in the signal path. The window, described elsewhere \cite{joint_3d-structured_2026}, is based on ultra-high-molecular-weight polyethylene (UHMWPE) and uses triangular grooves as a broadband anti-reflection structure. Since the inserted window is at room temperature and placed in front of the receiver, any additional loss contributes an effective input noise that can be compared with the change in the extracted $T_\mathrm{RF}$.

For each configuration, the same straight-line analysis was applied over the full set of LO and IF points, \corr{giving frequency-resolved estimates} $T_{\mathrm{RF,ref}}(\nu)$ and $T_{\mathrm{RF,win}}(\nu)$. The difference
\begin{equation}
    \Delta T_\mathrm{RF}(\nu)
    =T_{\mathrm{RF,win}}(\nu)-T_{\mathrm{RF,ref}}(\nu)
\end{equation}
represents the additional input noise introduced by the window. \corr{The same difference taken on the receiver noise $T_{noise}^{DSB}=T_{RF}+P^{\ast}/m$ is shown for comparison. The $\Delta T_{sys}$ gives a second independent transmission estimate, also shown in Fig.~\ref{fig:window_transmission}}. We convert these quantities into a physical transmission, using the relation:
\begin{equation}
    \tau(\nu)=
    \frac{T_{\mathrm{RF,ref}}(\nu)+T_\mathrm{phys}}
    {T_{\mathrm{RF,win}}(\nu)+T_\mathrm{phys}}
\end{equation}
where $T_\mathrm{phys}=295~\mathrm{K}$ is the physical temperature of the window. The transmission spectrum is compared with the FEM model of the grooved UHMWPE window. The agreement between the straight-line extraction and the simulated transmission gives a quality check of the calibration procedure and shows that the method can resolve small additional optical losses using only hot/cold receiver measurements performed in two configurations.

\section{Conclusion}

We have presented a numerical implementation of the straight-line method for extracting effective input-referred optical noise in a SIS heterodyne receiver. Using modern measurement setups, the method can be extended to a high-throughput analysis over a fine grid of LO frequencies, LO pump powers, and IF frequency bins. This overdetermined dataset allows the optical contribution to be reconstructed as a function of sky frequency, and also provides for consistency checks through the spread of crossing points, overlap regions, and pump-dependent stability measures.

For the receiver studied here, using the full pump set over the 4--16 GHz IF range, the median extracted effective input contribution was $T_{\mathrm{RF}}=33$ K, with a closure value of 4.1 K. Over the corresponding DSB receiver-noise analysis, the median measured $T_\mathrm{noise,DSB}$ was $57.0$ K. Although these numbers are not a noise-budget decomposition, they indicate that input losses represent a major contribution to the measured receiver noise. Reducing these losses, including warm optical losses, beam-efficiency loss, and spillover, is therefore a central challenge for approaching truly quantum-limited receiver performance. Attention is also required because the straight-line crossing extraction assumes that the mixer output noise varies weakly over the selected LO-pump range. Diagnostics on pump selection, crossing-point scatter, and overlap consistency are then essential to identify operating regions where non-constant $T_\mathrm{out}$ or mixer-related effects could lead to an overestimate of the optical loss.

\rev{The measurements were performed in DSB operation to demonstrate the numerical straight-line method using a single SIS mixer and IF amplification chain. This configuration avoids the additional complexity associated with operating two mixers and maintaining sideband separation over a wide IF bandwidth and varying LO pumping conditions. In DSB operation, each IF channel contains contributions from both the upper and lower sidebands. For the reconstruction of the sideband-resolved response, we assumed equal USB and LSB conversion gains at each IF frequency, although this assumption may not hold over a wide IF band. A 2SB receiver would provide a more direct approach by allowing the straight-line analysis to be performed independently on the USB and LSB outputs, which reduce the ambiguity in the assignment of optical noise to sky frequency and avoid the assumption of equal sideband gains. Such an implementation could improve the accuracy of the frequency-resolved characterisation. Although with a finite image rejection and additional IF-chain contributions plus variations in mixer balance with LO pumping, the method would need to be carefully calibrated.}

The numerical straight-line method provides a powerful practical tool for receiver characterisation. If combined with dense automated measurements and uncertainty quality quantification, it allows sensitive comparison of optical configurations, extraction of low-loss component transmission, and identification of the dominant optical limitations in wideband heterodyne receivers.

\section*{Acknowledgments}
The authors gratefully acknowledge the support received from the staff at the national facility for radio astronomy, the Onsala Space Observatory.

\bibliographystyle{IEEEtran}
\bibliography{references, ref2}

\end{document}